# Gated-Channel Conductivity Modulation by Hole Storage Effect Under Pulsed Conditions in *p*-GaN Gate Double Channel HEMT

*Hang Liao, Zheyang Zheng, Ji Shu, and Kevin J. Chen*

Department of Electronic and Computer Engineering, The Hong Kong University of Science and Technology, Clear Water Bay, Kowloon, Hong Kong, China

E-mail: hliaoac@ust.hk

**Abstract**

Recently, a *p*-GaN gate double channel HEMT (DC-HEMT) with conductivity modulation has been reported. The conductivity modulation is realized by hole storage in the gate stack and observed under quasi-static measurements. In this work, pulsed measurement and transient simulations of the DC-HEMT are carried out to disclose the conductivity modulation at high frequency and build-up time of hole storage. It takes 150 ns for the hole storage to be completely established despite a Schottky gate in the DC-HEMT with low gate leakage. The fast build-up of hole storage is attributed to the AlN insertion layer's strong confinement capability of holes and suppressed electron-hole recombination in the DC structure.

## 1. Introduction

GaN power devices have been intensively developed in recent decades and demonstrate desirable characteristics such as stable high-temperature operation along with favorable tradeoff between breakdown voltage and on-resistance.[1] This can be attributed to its superior material properties including a wide band gap, a high critical electric field (*E*-field), and high two-dimensional electron gas (2DEG) mobility. Nevertheless, GaN's direct bandgap is unfavorable to bipolar power devices such as bipolar junction transistors (BJTs) and insulated gate bipolar transistors (IGBTs), since the highly efficient electron-hole recombination leads to a short minority carrier lifetime in GaN, around 1 ns,[2] which is several orders of magnitude lower than that in the indirect bandgap semiconductors such as Si and SiC.[3], [4] This becomes a stumbling block to realize conductivity modulation which is beneficial to fully unlock the potential of GaN power devices.

Nonetheless, conductivity modulation has been observed in GaN PN diodes.[5] When the PN diode is forward biased, a vast number of electrons and holes would recombine to generate considerable photons due to the direct band gap nature of GaN. A portion of generated photons would in turn excite electron-hole pairs in doped GaN to realize conductivity modulation. Thus, conductivity modulation in GaN PN diodes is achieved by photon recycling. Yet, this approach is unsuitable for GaN HEMTs where the channel is undoped and electrons in the channel are polarization-induced.

*p*-GaN gate HEMT is the dominant architecture of GaN power devices. At forward gate bias, there is non-negligible hole injection from the gate electrode. The resultant co-existence of holes and electrons in the channel region satisfies the prerequisite for conductivity modulation. Based on this, a gate injection transistor (GIT) is proposed which features an ohmic gate and large gate leakage ($I_G$) to enable significant hole injection toward conductivity modulation.[6] Nevertheless, the short lifetime of the minority carrier (hole) and the high vertical *E*-field in the gate stack which would sweep injected holes away from the channel collectively make conductivity modulation relatively weak in this device.

Recently, a normally-off *p*-GaN Schottky gate double channel high-electron-mobility transistor (DC-HEMT) has been demonstrated on a commercialized 6-inch GaN-on-Si wafer.[7] It features a delicate AlN insertion layer (AlN-ISL) which spatially separates electrons and holes and thus the minority carrier (hole) lifetime is prolonged. Besides, the AlN-ISL confines the injected holes in the vicinity of the channels. As a result, hole storage occurs in the gate stack to induce more electrons in channels. Consequently, conductivity modulation is realized in the DC-HEMT evidenced by a distinct hole-storage-induced second gate modulation ($g_m$) peak and 15% larger saturation current ($I_{DSAT}$) than a conventional *p*-GaN gate single channel (SC) HEMT.[7]

Notably, these results are obtained under quasi-static measurements. A Schottky gate is adopted in the DC-HEMT for compatibility with the mainstream *p*-GaN gate technology and voltage-driven scheme. Considering the low $I_G$ of a Schottky gate, it would take time for the hole storage to build up, and thus the conductivity modulation might be absent under high-frequency operations. Hence, it is essential to characterize the DC-HEMT under the pulsed mode to mimic the fast-switching condition in real applications.

In this work, pulsed measurements and transient simulations are done to reveal the build-up time of hole storage. The hole storage can build up in 150 ns, suggesting the DC-HEMT can offer conductivity modulation in operation up to 1 MHz. Such fast build-up of hole storage relies on the strong hole confinement capability of the AlN-ISL and mitigated electron-hole recombination enabled by the DC structure.

## 2. Device Structure

The fabrication process of the 200-V DC-HEMTs is described in our previous work.[8] Devices under test (DUTs) in this work feature a gate-to-drain length ($L_{GD}$) of 6 μm, a gate length ($L_G$) of 4 μm, a gate-to-source length ($L_{GS}$) of 2 μm, and a gate width ($W_G$) of 20 μm. The cross-sectional schematic of the DC-HEMT is shown in **Figure 1a** and the band diagram along the A−A' cutline at positive gate voltage ($V_{GS}$) is depicted in **Figure 1b**.

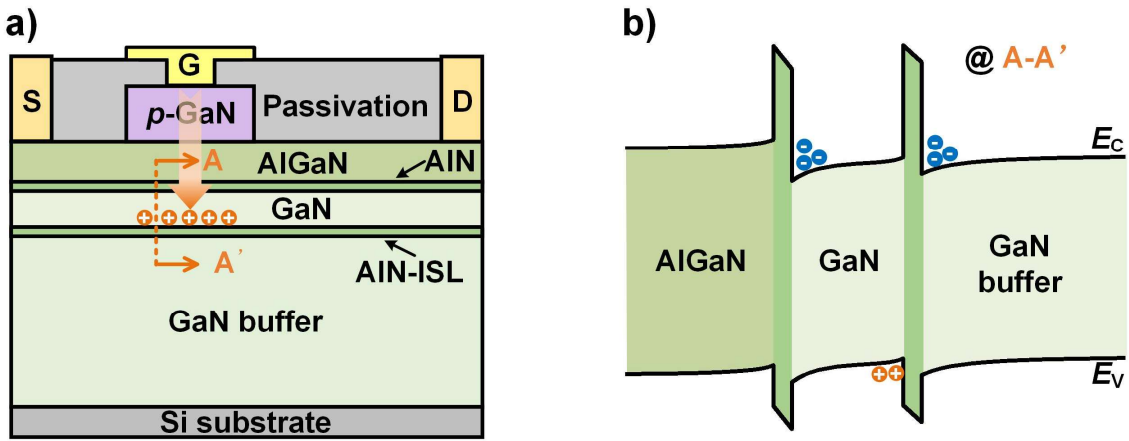

**Figure 1.** a) Schematic cross-section of the DC-HEMT and b) band diagram in the gated region.

The AlN-ISL effectively confines the injected holes from the gate electrode at a positive $V_{GS}$ to enable hole storage at the lower interface of the upper GaN layer (**Figure 1**). These holes are spatially separated from the electrons in the channels and thus the recombination process is weakened (**Figure 1b**). Yet, they are sufficiently close such that the confined holes could effectively modulate the electron density in the channels.

## 3. Characterizations

### 3.1. Comparison of Pulsed and Quasi-static Measurement Results in DC-HEMT

The pulsed transfer measurement is carried out with a pulse width of 500 ns which is limited by the equipment. The pulse separation is set to 10 ms to eliminate the impact from preceding pulses (**Figure 2a**). The quasi-static measurement is also conducted, and the schematic waveform is shown in **Figure 2b**. The difference between these two measurement conditions lies in the intermittent or the constant hole injection as indicated in **Figure 2a** and **Figure 2b**. The intermittent hole injection is potentially challenging to the establishment of hole storage in the pulsed condition.

According to the pulsed transfer measurement result (**Figure 2c**), two clear $g_m$ peaks are still found. The first $g_m$ peak is induced by the fully-on of the lower channel. The second $g_m$ peak which is induced by hole storage takes place at the same $V_{GS}$ (6.5 V) and features the same amplitude of $g_m$ (80 mS/mm) as the quasi-static result. Besides, the pulsed measurement result yields the same current level as the quasi-static one. The steep increase in $I_D$ at $V_{GS}$ = 2 V and 6 V is corresponding to two $g_m$ peaks.

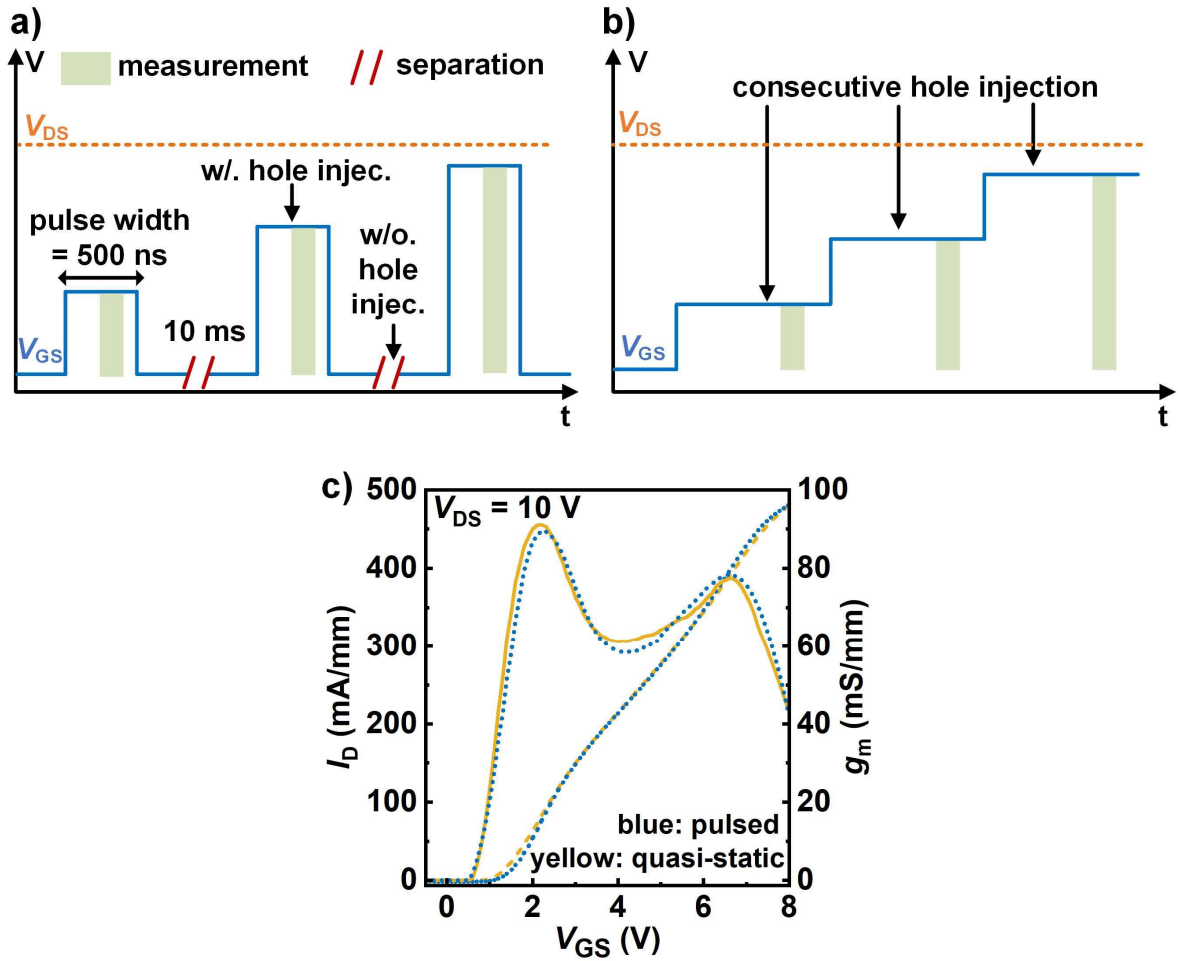

**Figure 2.** Schematic waveforms of a) the pulsed and b) quasi-static transfer measurements. c) Pulsed and quasi-static transfer characteristics of the DC-HEMT at $V_{DS}$ = 10 V.

Output characteristics are also obtained in both conditions. The schematic pulsed and quasi-static waveforms for the measurements are shown in **Figure 3a** and **Figure 3b**, respectively. These two scenarios show similar on-resistance ~7 Ω·mm and saturation current level ~500 mA/mm, indicating negligible trapping states in the DC-HEMT.

In the pulsed result (**Figure 3c**), the strong gate modulation in the saturation region (with a notable gap between two adjacent lines) at high $V_{GS}$ is observed, which corresponds to the second $g_m$ peak observed in the pulsed transfer characteristics at $V_{GS}$ = 6.5 V (**Figure 2c**). The self-heating-induced drain current decrease is alleviated in pulsed output measurement results.

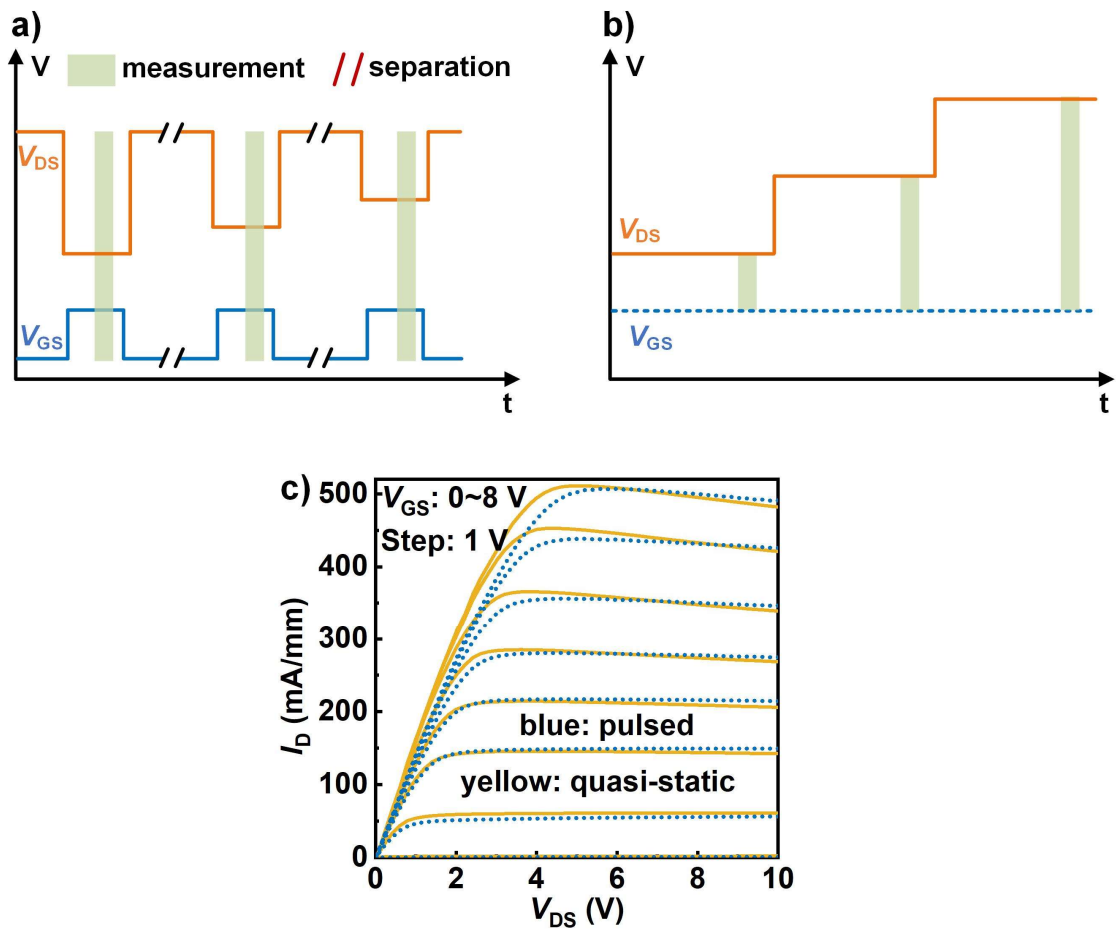

**Figure 3.** Schematic waveforms of a) the pulsed and b) quasi-static output measurement. c) Pulsed and quasi-static output characteristics of the DC-HEMT.

These results suggest that the hole storage can be established within 500 ns with the same density as the quasi-static condition. Considering the hole storage originates from the $I_G$,[9] the fast build-up of hole storage under the pulsed condition might be induced by a larger $I_G$ than the quasi-static counterpart. To verify this, the pulsed and quasi-static gate performances are characterized with drain and source electrodes grounded as shown in **Figure 4a** and **Figure 4b**. The insets show the schematic waveforms of $V_{GS}$.

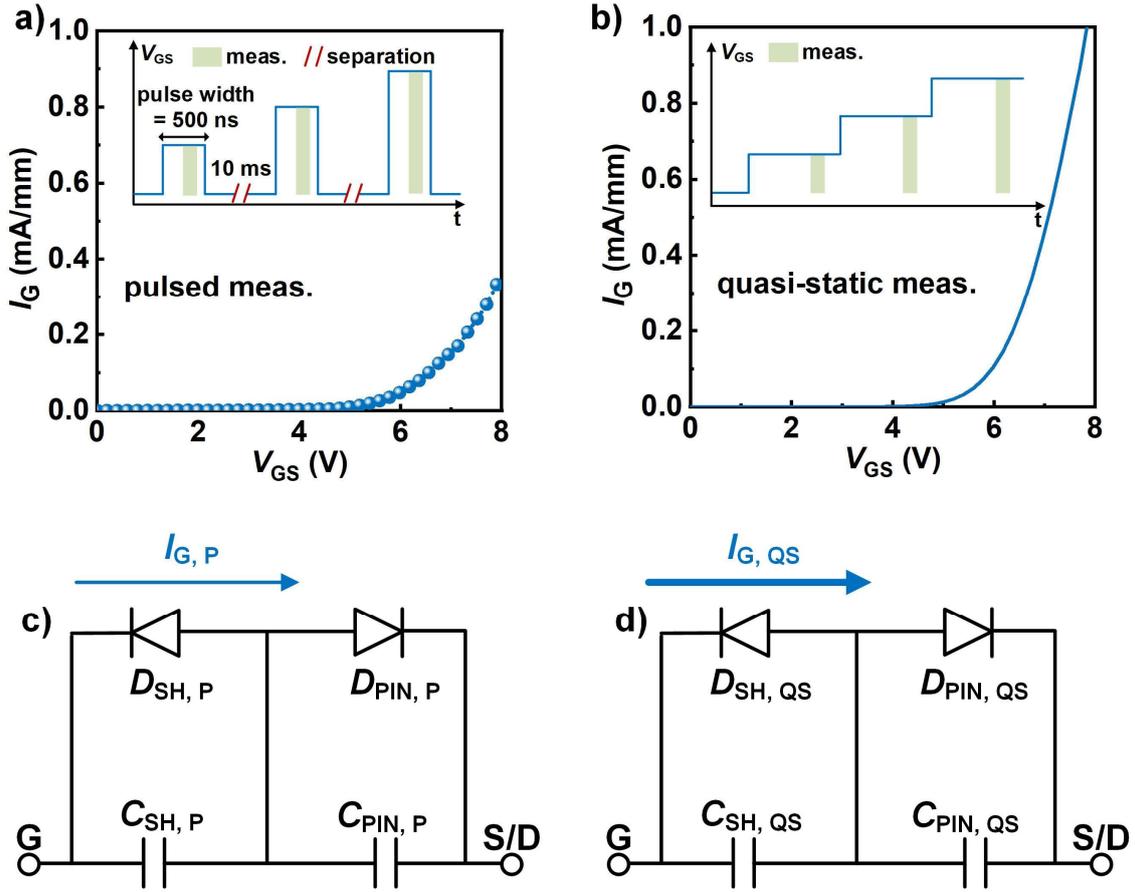

**Figure 4.** Gate characteristics under a) the pulsed measurement and b) the quasi-static measurement. Insets: schematic waveforms. Equivalent circuit model of the two junctions in the gate stack c) under pulsed condition, and d) under quasi-static condition.

According to the measurement results, the pulsed $I_G$ is lower than the quasi-static one. This can be explained by the equivalent circuit model of two back-to-back diodes (Schottky diode $D_{SH}$ and *p-i-n* diode $D_{PIN}$) in the gate stack as shown in **Figure 4c** and **Figure 4d**. [10], [11] In the pulsed mode, the *p*-GaN depletion region cannot be formed immediately, leading to a smaller depletion region width than in the quasi-static mode. Thus, the Schottky junction capacitor would exhibit a higher capacitance in the pulsed mode ($C_{SH, P}$) than the quasi-static one ($C_{SH, QS}$). According to the voltage partition between two series capacitors, the voltage drop in $C_{SH, P}$ (and $D_{SH, P}$) is smaller compared with $C_{SH, QS}$ (and $D_{SH, QS}$). In the DC-HEMT, the $I_G$ is dominated by the hole current since the electron spill-over in the lower and upper channel is suppressed by the additional AlN-ISL barrier and hole storage, respectively. [12], [13] As a result, $I_G$ decreases with a lower voltage partition in the $D_{SH, P}$ under the pulsed mode. These results rule out the possibility that the fast-build-up hole storage under the pulsed condition is caused by a larger $I_G$. Instead, it is realized by the high efficiency of utilizing injected holes from the $I_G$, which depends on the AlN-ISL's strong confinement capability.

## 4. Simulations and Discussions

The pulsed measurement results prove the hole storage can completely build up within 500 ns while the specific value is unknown. To figure it out, transient simulations are carried out. The schematic $V_{GS}$ waveform in transient simulations is shown in **Figure 5a**. The rising edge is set to be 10 ns to minimize its impact on hole storage. During on-state, the density of hole storage is read out every 10 ns. Other settings in simulations are summarized in **Table 1**. The thickness of each layer in the epitaxy is corresponding to the DUT.

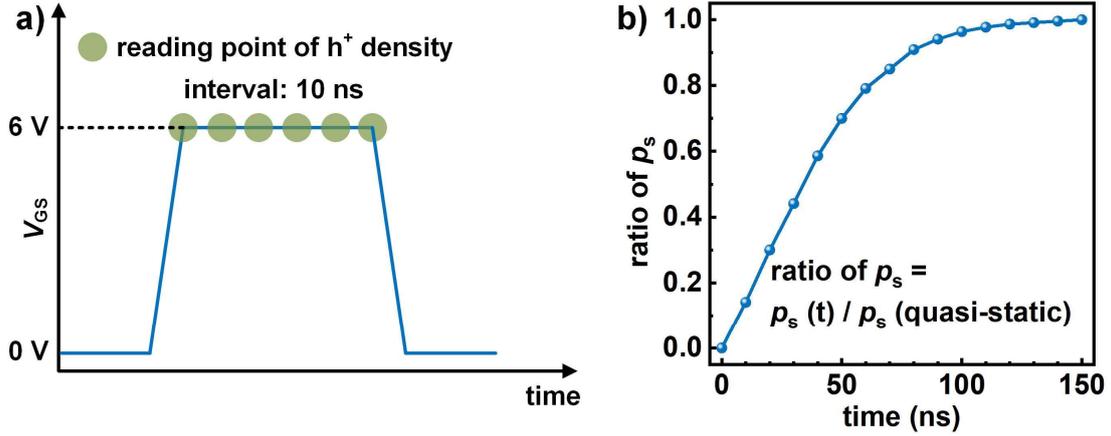

**Figure 5.** a) Schematic $V_{GS}$ waveform in the transient simulation with a rising edge of 10 ns. b) Relation between ratio of $p_s$ and time.

Table 1. Simulation settings.

| Parameters | Value | Unit |
| --- | --- | --- |
| Pulse rising edge | 10 | ns |
| Radiative recomb. coefficient | $1.1 \times 10^{-8}$ | $cm^3s^{-1}$ |
| Gate Schottky barrier height | 0.6 | eV |
| $p$-GaN thickness | 100 | nm |
| Activated Mg concentration | $3 \times 10^{18}$ | $cm^{-3}$ |
| Activation energy of Mg in GaN | 186 | meV |
| $Al_{0.2}Ga_{0.8}N$ thickness | 13.5 | nm |
| AlN thickness | 1 | nm |
| Upper GaN layer thickness | 6 | nm |
| AlN-ISL thickness | 1 | nm |
| Lower GaN layer thickness | 240 | nm |
| Buffer layer thickness | 2.6 | μm |

According to the simulation results in **Figure 5 (b)**, it takes 150 ns for the hole storage to fully build up with the same density as that at quasi-static state. The fast hole storage is attributed to two factors: 1) high-efficiency utilization of gate leakage where injected holes are well-confined by the AlN-ISL; 2) weakened recombination by delicate channel design where stored holes and electrons are spatially separated.

AlN-ISL's confinement capability of holes is determined by its thickness ($t_{AlN}$). With a thicker AlN-ISL, the thermal emission and tunnelling of holes would both be suppressed and thus it can yield a higher density of hole storage.

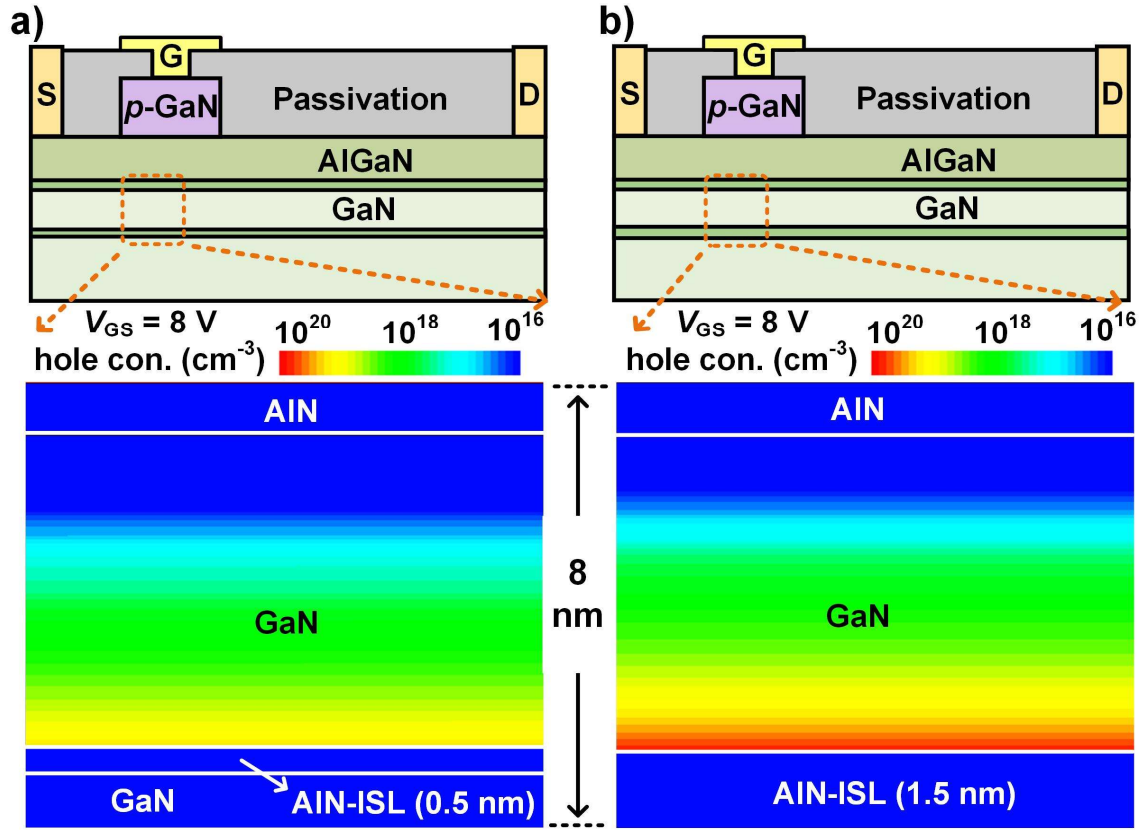

**Figure 6.** The simulated density of hole storage at $V_{GS}$ = 8 V with a) $t_{AlN}$ = 0.5 nm, and b) $t_{AlN}$ = 1.5 nm. c) The simulated hole storage density at different $V_{GS}$.

In the simulation result, there is a significant difference in hole storage concentration (at the lower interface of the upper channel layer) with $t_{AlN}$ = 0.5 nm and $t_{AlN}$ = 1.5 nm (**Figure 6a and 6b**). The peak hole concentration in $t_{AlN}$ = 1.5 nm is one order of magnitude higher than $t_{AlN}$ = 0.5 nm. This result indicates that the AlN-ISL's confinement capability of holes is highly sensitive to its thickness due to the strong polarization field within this layer.

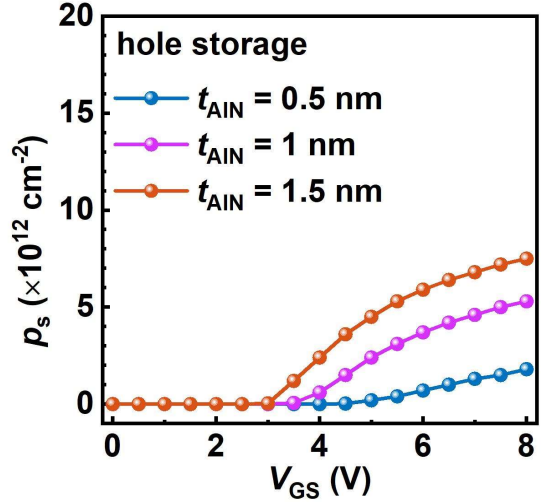

**Figure 7.** The simulated dependence of hole storage density on $V_{GS}$ with different $t_{AlN}$.

The relation between the density of hole storage and $V_{GS}$ with different $t_{AlN}$ is quantified and shown in **Figure 7**. With $t_{AlN} = 0.5$ nm, the hole storage is insignificant even at high $V_{GS}$ since injected holes are swept to cross the ultra-thin AlN-ISL by the high $E$-field in the gate stack. In this case, the carrier dynamics is similar to that in the SC-HEMT where injected holes are swept away from the channel and toward the buffer layer promptly. [14] The hole storage occurs at a lower $V_{GS}$ with a thicker AlN-ISL, indicating a moderate-level hole injection could also lead to hole storage if injected holes can be well-confined. For scenarios with non-negligible hole storage ($t_{AlN} = 1$, 1.5 nm), the increase in hole density ($p_s$) steadily slows down as $V_{GS}$ increases. This is caused by the radiative recombination between electrons and holes. The radiative recombination rate can be obtained by:

$$R = B \times n \times p$$

where $B$ is the radiative recombination coefficient in the unit of cm$^3$s$^{-1}$ and it is determined by the material properties. $n$ and $p$ represent the hole and electron density ($n_s$), respectively. As $V_{GS}$ increases, both $p_s$ and $n_s$ increase accordingly. Consequently, the increase rate of $p_s$ will be limited by the radiative recombination.

In the delicate-designed DC structure, the separation of electrons and holes effectively suppresses the radiative recombination by staggering electrons' and holes' concentration peaks as depicted in **Figure 8(a)** with $t_{AlN} = 1$ nm. The corresponding radiative recombination rate in both channels is shown in **Figure 8(b)**. The radiative recombination is significantly mitigated in the lower channel due to a much lower hole density there compared to the upper one. This is reached with the strong confinement capability of the AlN-ISL.

Theoretically, the $p_s$ can be further boosted with a thicker AlN-ISL while this would lead to a lower threshold voltage ($V_{TH}$). To obtain a desired trade-off between hole storage density and $V_{TH}$, $t_{AlN}$ = 1 nm is opted in the DC-HEMT. For further optimization of the DC-HEMT, there is an approach to solve the dilemma toward a higher density of hole storage without lowering $V_{TH}$. The thickness of the AlGaN barrier layer can be further reduced and the equivalent $t_{AlN}$ can be accordingly increased to keep the $V_{TH}$ same. The side effect of the scenario could be stronger electron and hole spillover in the gate stack, leading to a larger $I_G$. However, the $I_G$ in the DC-HEMT is two orders of magnitude lower than the conventional SC-HEMT. Thus, there is still room to engineer the AlGaN layer and AlN-ISL toward more promising device performance of the DC-HEMT. Furthermore, the DC structure is expected to lay a solid foundation for the development of novel device architectures. [15], [16]

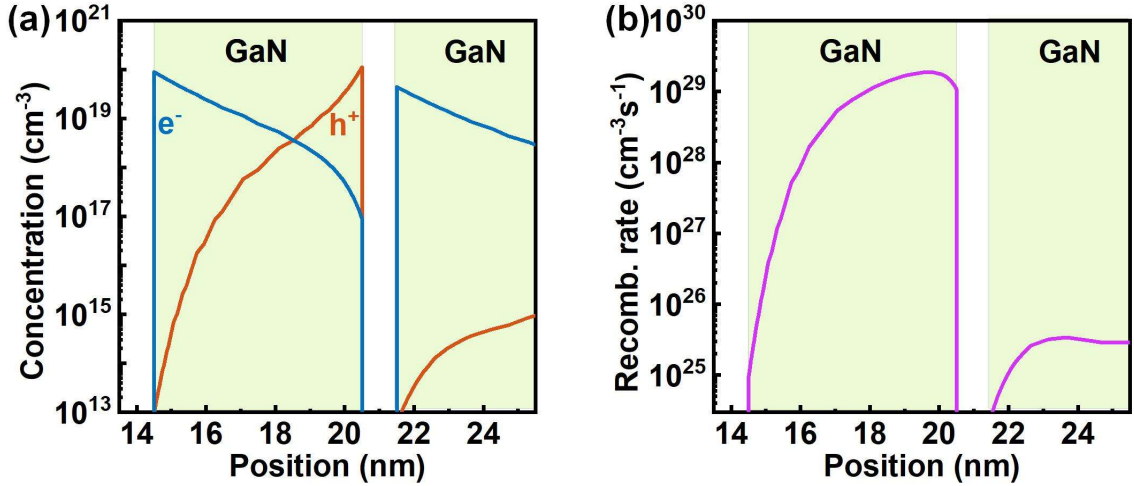

**Figure 8.** a) The simulated hole and electron distribution in both channels. b) Radiative recombination rate in both channels with $t_{AlN}$ = 1 nm.

## 5. Conclusion

A *p*-GaN gate DC-HEMT is demonstrated with conductivity modulation under both quasi-static and pulsed conditions. Despite the low $I_G$ in the DC-HEMT, the hole storage can still build up in 150 ns, since the AlN-ISL features strong capability of confining holes and the recombination is suppressed. The AlN-ISL's confinement ability is quantified by the simulation where the density of hole storage shows a strong dependence on $t_{AlN}$ while the alleviated recombination is achieved by staggering electron and hole concentration peaks enabled by the quantum-well DC structure. These findings serve as guidance for understanding the physics in the DC-HEMT and further optimizing its performance.